\documentclass{article}
\usepackage[utf8]{inputenc}
\usepackage[dvipsnames,table,xcdraw]{xcolor}
\usepackage{framed} % shaded* environment
\usepackage{graphicx}
\usepackage{graphics}
\usepackage{subcaption}
\usepackage[margin=1in]{geometry}
\graphicspath{{figs/}}

\usepackage{cliff}

\usepackage[breaklinks=true,bookmarks=false]{hyperref}
\hypersetup{
  colorlinks=false,%
  pdfborder = 0 0 0
}
\definecolor{pastelblue}{rgb}{0.68, 0.78, 0.81}

\colorlet{shadecolor}{pastelblue}

\blockpar

\title{Talking to GDELT Through Knowledge Graphs}
\author{Audun Myers, Max Vargas, Sinan G. Aksoy, Cliff Joslyn, Benjamin Wilson, \\ Lee Burke, Tom Grimes}

\date{June 2, 2025}
\date{}

\begin{document}

%\begin{itemize}
%    \item Cliff: will create responses to reviewers, but will ask if we need/can.
%    \item Audun: going to rework language to address comments.
%    \item Cliff: going to write about the structure of the KG vs standard ontology from table (not a simple star).
%    \item Audun (Done): will create evaluation (quantitative) results
    %\item Max: will create more questions.
%    \item unassigned: Create table of statistics between graphs in figure 3.
%\end{itemize}

\maketitle

%IS EVERYONE OKAY WITH ADDING IN THEIR EMAILS on the final publication (had to add them for submission)?
%{\tiny \color{red}
%TO DO:

%\begin{enumerate}
%    \item Make an emphasis on we are still working on the %LlamaIndex route.
%    \item Make sure to include an example where it (GDELT %KG) did well? - Probably by using the grph query results.
%    \item Do a basic graph query result.
%    \item Do an experiment testing different families of %questions and how each pipeline performs. is there a method %that performs best with certain types of questions?
%    \item Make an argument that a graph query works for %GDELT KG but not LlamaIndex KG because it doesnt follow the %structure properly.
%    \item Then we only use g-retriever on GDELT KG because %we already have it so why not?
%    \item Look for standard package for standard graph %queries.
%    \item Need to do specific GDELT KG lit review - mention %that the standard GDELT analysis is really basic - ADD %LINKS.
%    \item Assign writing tasks to everyone.
%    \item add lit review of the "talking to graph" crowd.
%    \item Ask question to LLM: describe the main events
%    \item repeat experiments with reduced ontology

%\end{enumerate}
%}

%\tableofcontents

\begin{abstract}
In this work we study various Retrieval Augmented Regeneration (RAG) approaches to gain an understanding of the strengths and weaknesses of each approach in a question-answering analysis.
To gain this understanding we use a case-study subset of the Global Database of Events, Language, and Tone (GDELT) dataset as well as a corpus of raw text scraped from the online news articles.
To retrieve information from the text corpus we implement a traditional vector store RAG as well as state-of-the-art large language model (LLM) based approaches for automatically constructing KGs and retrieving the relevant subgraphs. In addition to these corpus approaches, we develop a novel ontology-based framework for constructing knowledge graphs (KGs) from GDELT directly which leverages the underlying schema of GDELT to create structured representations of global events. 
For retrieving relevant information from the ontology-based KGs  we implement both direct graph queries and state-of-the-art graph retrieval approaches.
We compare the performance of each method in a question-answering task. We find that while our ontology-based KGs are valuable for question-answering, automated extraction of the relevant subgraphs is challenging. Conversely, LLM-generated KGs, while capturing event summaries, often lack consistency and interpretability. Our findings suggest benefits of a synergistic approach between ontology and LLM-based KG construction, with proposed avenues toward that end. 

\end{abstract}
%!TEX root = ..\main.tex
%-------------------------------
%*******************************

\section{Introduction}
%\TGComment{this is an example comment}

In this work we study several approaches for communicating with a corpus of text via relevant text and knowledge graph (KG) representation and retrieval facilitated by Large Language Models (LLMs). Our goal is to understand the benefits and drawbacks of  Retrieval Augmented Generation (RAG) approaches to  corpus management and anlysis  when combined with an LLM. Throughout we use as a case study a novel KG derived from the Global Data on Events, Location, and Tone (GDELT)\footnote{https://www.gdeltproject.org/}~\cite{leetaru2013gdelt} dataset. 

%We also demonstrate how to leverage a LLM to interact with these KGs to gain an understanding of their usefulness.

%\subsection{Literature for Talking to Graphs}
As a way to enhance LLM outputs, researchers and practitioners have been quick in applying LLMs to query and understand proprietary data through retrieval-augmented-generation (RAG) \cite{10.5555/3495724.3496517}. It has been shown that reasoning over the typical RAG framework, which only takes advantage of the unstructured text articles, fails to capture global information about the provided data \cite{edge2024localglobalgraphrag, Xu_2024}. 

Motivated by this limitation,  there has been recent interest in adapting these techniques to the case where our data has a graph structure, so that the LLM can directly ingest important relationships in the knowledge base \cite{edge2024localglobalgraphrag, he2024g, mavromatis2024gnnraggraphneuralretrieval, zhu2023anetscalablepathbasedreasoning}. More specifically, KGs \cite{HoABlE21} are graph structures which are richly attributed with typing and semantic information on both nodes and edges. KG techniques provide ways to automatically query and extract information stored in a KG without the user explicitly needing to understand query languages to probe their knowledge base. Typically, these AI-based search algorithms find subgraphs that can be used to answer a user-provided query. 

The interactions between KGs and LLMs have potential beyond merely question-answering and knowledge extraction (see different research directions outlined by Pan et al. \cite{Pan_2024}). In particular, reflective of KGs being used to enhance LLM outputs, LLMs can be used to enhance existing KGs or create new ones entirely from scratch. However, exploration of techniques to this end either (1) do not deal with imposing different ontological structures in graph creation or (2) only focus on extracting ontological structures using LLMs \cite{trajanoska2023enhancingknowledgegraphconstruction,yao2024exploringlargelanguagemodels}. 

%\subsection{GDELT introduction and literature review}
Throughout this work we use the GDELT dataset as a case study. GDELT is a massive collection of news reports that provide a real-time computational record of global events that is published every 15 minutes. It aggregates information from various news sources, blogs, and social media platforms to construct a large collection of data including information on people, organizations, locations, themes, and emotions.
Essentially, GDELT offers a snapshot of the world's collective events, enabling researchers and analysts to explore complex patterns and relationships within global society. By analyzing this data, it's possible to identify emerging trends, assess risks, understand public sentiment, and track the evolution of various issues over time. 
The applications of GDELT are diverse and far-reaching. Some of the most common use cases including event monitoring \cite{owuor2023temporal, owuor2020tracking, yonamine2013nuanced}, risk assessment and prediction \cite{galla2018predicting, qiao2016predicting, qiao2017predicting, voukelatou2020estimating, wu2017forecasting, yonamine2013predicting}, and social science research \cite{alamro2019predicting, bodas2016using, boudemagh2017news, keertipati2014multi}.

GDELT describes its structure as a Global Knowledge Graph (GKG, specifically, we use the the Global Knowledge Graph edition 2 (GKG2) of GDELT). But in fact GDELT-GKG2 is implemented as multiple linked tables recording information about the relationship between articles and events, and thus effectively has the structure of a relational database.
%(see more details in section~\ref{ssec:gdelt}). 
Another important contribution of this paper is to actually realize GKG2 properly in the mathematical form of a KG, effectively a graph database, derived from and consistent with its native relational database form. 
%they do not provide a clear methodology for doing this. 
To facilitate this effort, we have identified a lightweight ontology for GDELT in the form of its graph schema, realizing its relational database schema in a KG form.

Using the KG that we construct from the GDELT-GKG2 dataset, we provide a case study to explore the utility of LLM-based tools to extract information and confirm that the KG can be used for question-answering in cases where traditional RAG fails. As part of our analysis, we compare to KGs produced from processing various news articles with an LLM, prompting it to try and adhere to a reduced version of the same ontology.

\begin{figure}[h!]
    \centering
    \includegraphics[width=0.85\textwidth]{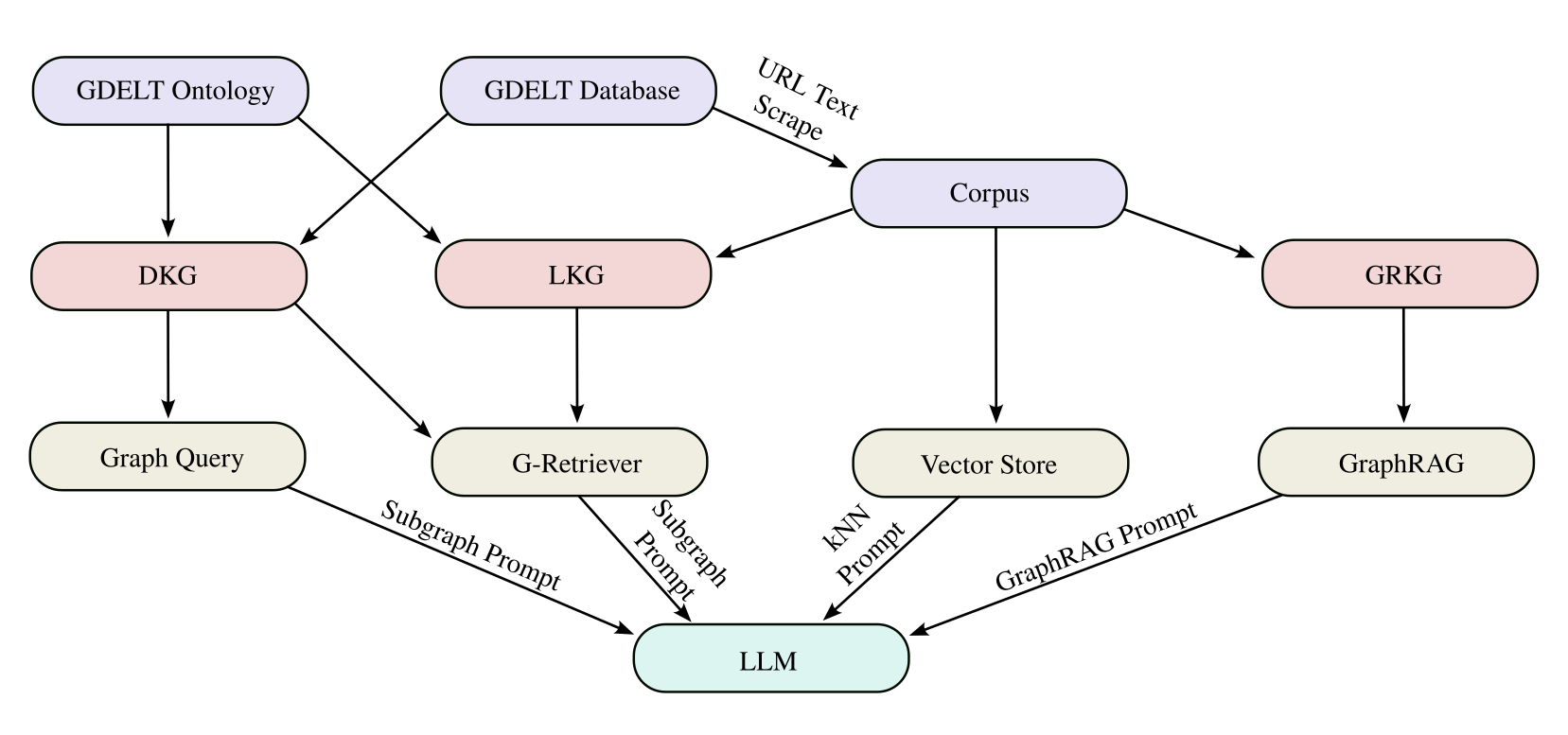}
    \caption{Pipeline of different experiments ran to analyze the GDELT database using an LLM.}
    \label{fig:experiments_pipeline}
\end{figure}

The current state of neurosymbolic work is noted for the plethora of experimental architectures available. While  details are  explicated below in Section \ref{sec:kgmeth}, we  preview ours in Figure \ref{fig:experiments_pipeline}, including the five methodological pathways which are quantitatively compared: 1) graph queries on the KG (called the DKG) derived ``directly'' from GKG2; 2) use of G-Retriever\footnote{https://github.com/XiaoxinHe/G-Retriever} \cite{he2024g} against the same DKG; 3) RAG against a vector store representation of GKG2; 4) G-Retriever against a second KG (called the LKG) derived from using Llamaindex\footnote{https://www.llamaindex.ai/} \cite{LLamaIdxKG} against the GDELT source articles; and 5) GraphRAG\footnote{https://microsoft.github.io/graphrag/} Q\&A deployed against a third KG (called GRKG) using Microsoft's open-source GraphRAG package with default configuration parameters. 

\section{Constructing a Knowledge Graph for GDELT}

As previously mentioned, while the GDELT-GKG2 dataset is not actually natively in the form of a knowledge graph, it is advertised and frequently cited as being one. We believe that we are making a distinct contribution to the research community by converting the very popular GKG2 database into a proper KG.

GKG2 is natively a database of three related tables:
\begin{itemize}
    \item {\tt expert.csv} captures event information;
    \item {\tt GKG.csv} captures article information; and 
    \item {\tt mentions.csv} relates which articles mention which events. 
\end{itemize}
%Here we propose an ontology based on the database schema to construct a knowledge graph to be analyzed using an LLM.

%\subsection{GDELT Database Schema} \label{ssec:gdelt}

\begin{figure}[h!]
    \centering
    \includegraphics[width=0.95\textwidth]{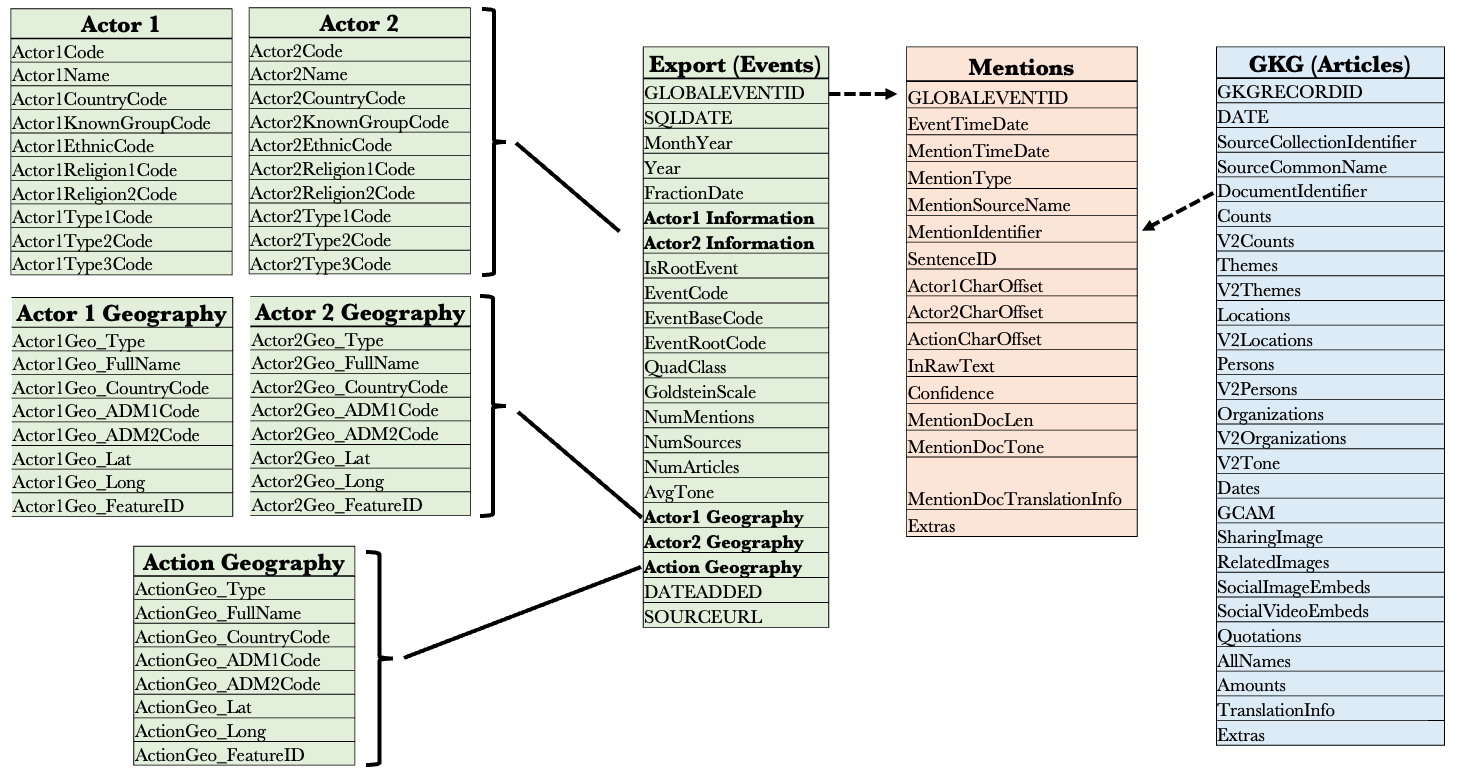}
    \caption{GDELT GKG 2.0 schema relating articles (GKG), mentions, and events (Export).}
    \label{fig:gdelt_schema}
\end{figure}

The database schema for these three CSV files  is shown in Fig.~\ref{fig:gdelt_schema} (see also \cite{jayanetti2023exploring}). The key characteristics of this database relational schema should be interpreted as follows:
\begin{itemize}
    \item The three tables are color-coded by Events (green), Mentions (pink), and Articles (blue). 
    \item The single table of Events is shown in multiple green sub-tables, simply for clarity and convenience to layout a long record structure.
    \item Single-headed arrows represent one-to-many relationships between the tables. Specifically:
    \begin{itemize}
        \item Each Event maps to multiple Mentions via the shared \verb+GLOBALEVENTID+ field.
        \item Each Article maps to multiple Mentions via the \verb+DocumentIdentifer+ field on the Article side matching to the \verb+MentionIdentifier+ field on the Mention side.
    \end{itemize}
    
    \item In this way, the Mentions table acts as a ``pairs file'' recording a many-many relation between Events and Articles: each event can be  mentioned in multiple articles, and dually each article can mention many events.
    Each Article also has both a unique identifier through the \verb+GKGRECORDID+ or the \verb+DocumentIdentifer+ fields, since each row in the GKG data represents a single article.
    %These relations between the three tables were also, separately, found in~.
\end{itemize}

%We additionally have no-headed arrows which are used for visualization purposes only. Specifically, we expand the multiple fields in the Export data with the Actors and Action information such as geography and name, and other types for each. 
%\subsection{Ontology for GDELT}

\begin{figure}[h!]
    \centering
    \includegraphics[width=0.99\textwidth]{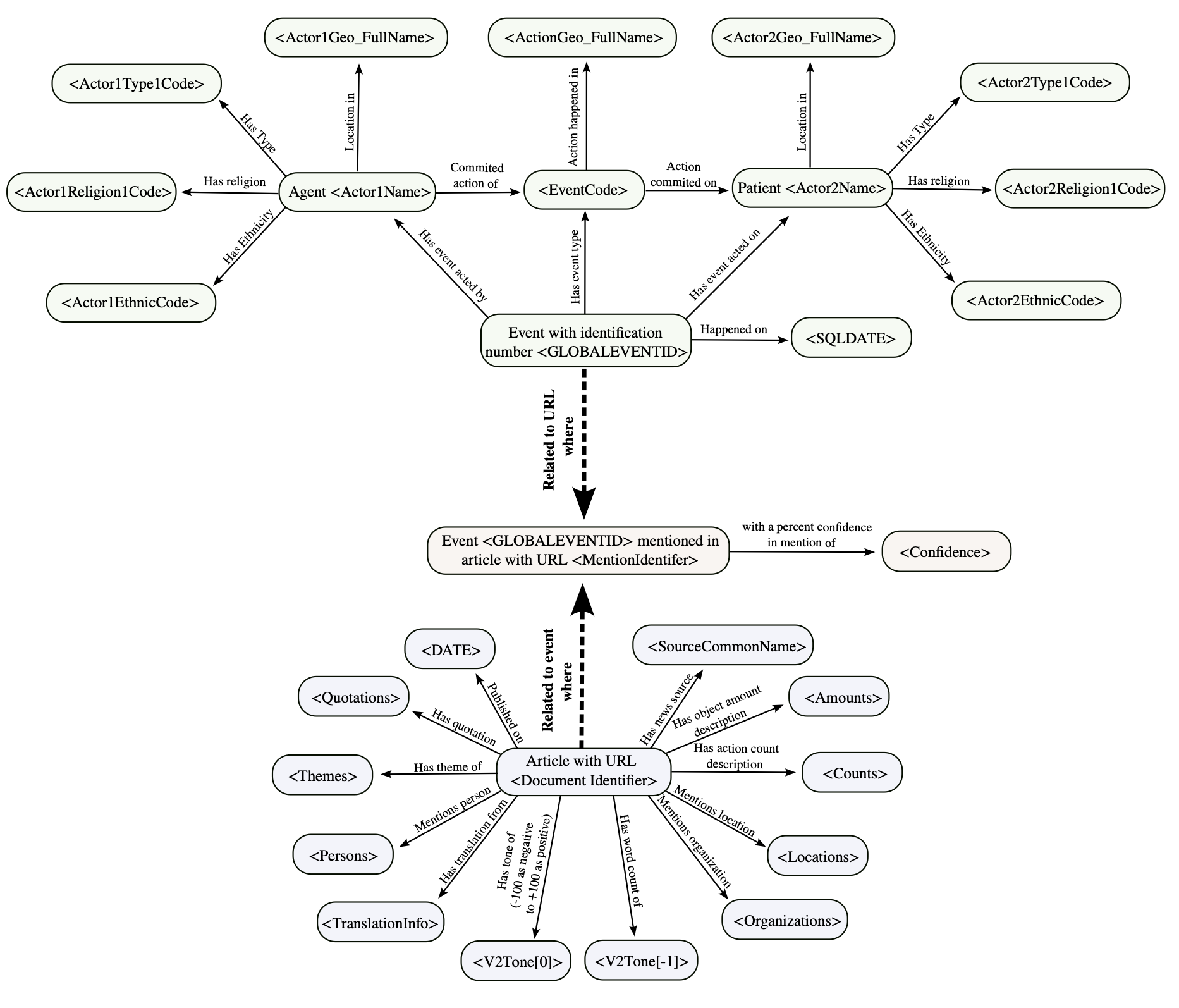}
    \caption{GDELT GKG 2.0 ontology relating articles and events..}
    \label{fig:gdelt2_ontology_draft}
\end{figure}

Methods to automatically determine the graphical form of a relational database are widely known \cite{SeJTiS11}. Most naturally, consider a table $T$ with $m$ rows $T[j], 1 \le j \le m$ and $n$ columns $T.i, 1\le i \le n$. Then each of the $m$ rows $T[j]$ is represented as a node in one meta-class labeled by the primary keys. This node then has $n$ outgoing edges, each connecting to a node in another meta-class representing the field value $T[j].i$, and labeled by the column name. The resulting ``star'' bipartite graphs are then linked over shared values, including across multiple tables. This method straightforwardly produces a graph schema consistent with a given RDB, which may or may not be of sufficient complexity to warrant the lofty description of ``ontology''. In our case, such a straightforward approach is mostly warranted, although as we will see additional constraints in the event table will argue for a somewhat more specific and idosyncratic graph structure.

After understanding the GDELT database schema, we developed a capability to convert (portions of) the GDELT database to a KG using an ontology as a graph typing schema, derived from the above relational schema. This is shown in Fig.~\ref{fig:gdelt2_ontology_draft}, to be interpreted as follows:
\begin{itemize}
    \item Nodes in the ontology indicate the types of nodes possible in the KG.
    \item Nodes are color-coded to indicate their source relational table.
    \item Fields in $\langle \mbox{angled brackets} \rangle$ indicate the field name in the schema.
    \item Solid edges indicate a field in a relational table and are labeled with the type of semantic relation.
    \item Dashed and bold edges indicate the structural, one-to-many relations in the relational schema. 
\end{itemize}
%This ontology is color coded by the three csv files provided: Events-export.csv (green), Mentions-mentions.csv (orange), and Articles-gkg.csv (blue). The dashed arrows represent the relations between the csv files which are one-to-many relations with an article mentioning many events, and an event being mentioned by many articles. 
The naming convention also captures the unique identifier for these csv files, so that $\tup{\h{\tt GLOBALEVENTID}}$ identifies unique Events, the pair $(\tup{\h{\tt GLOBALEVENTID}}, \tup{\h{\tt MentionIdentifier}})$ identifies unique Mentions, as does $\tup{\h{\tt DocumentIdentifier}}$ for Articles. We again note that the document and mention identifiers are the same field, but have different field names (e.g., a URL is typically used for the document identifier and the same URL is used for the mention identifier).

%One potential issue when using the full ontology provided here is the size of the graph. As such, it may be more prudent to remove node and edge types from the graph to work with a more manageable-sized knowledge graph. However, for the remainder of this work we will use the full ontology for our analysis.

\section{Case Study - Baltimore Bridge Collapse}

Here we will provide an analysis of data collected over a recent and short period of time to do question-answering based analysis. The point of collecting recent data is that the LLMs used have not yet been trained on these events (at the time of this study) and thus the knowledge systems are needed to supply the LLM with relevant information. Specifically, this analysis uses a subset of the GDELT data collected on March 26th of 2024 from 12:00 AM to 10:00 AM during and after the collapse of the Francis Scott Key Bridge in Baltimore, Maryland, which occurred at approximately 1:29 AM. This 10 hour window of time captures the media response to this disaster. We filtered down the collected mentions data to only include rows in any of the related data if it included any of the keywords ``Baltimore'', ``bridge'', ``collapse'', or ``ship''. We then used all \+GLOBALEVENTIDs+ and \+MentionIdentifiers+ in this reduced mentions file to collect the relevant events and articles. This filtration resulted in using approximately 1.33\% of the available data with 371 events, 2047 mentions, and 209 articles. 

\subsection{GDELT Knowledge Graphs}
\begin{figure}[h!]
     \centering
     \begin{subfigure}[t]{.31\textwidth}
         \centering
         \includegraphics[angle=90,width=0.99\textwidth]{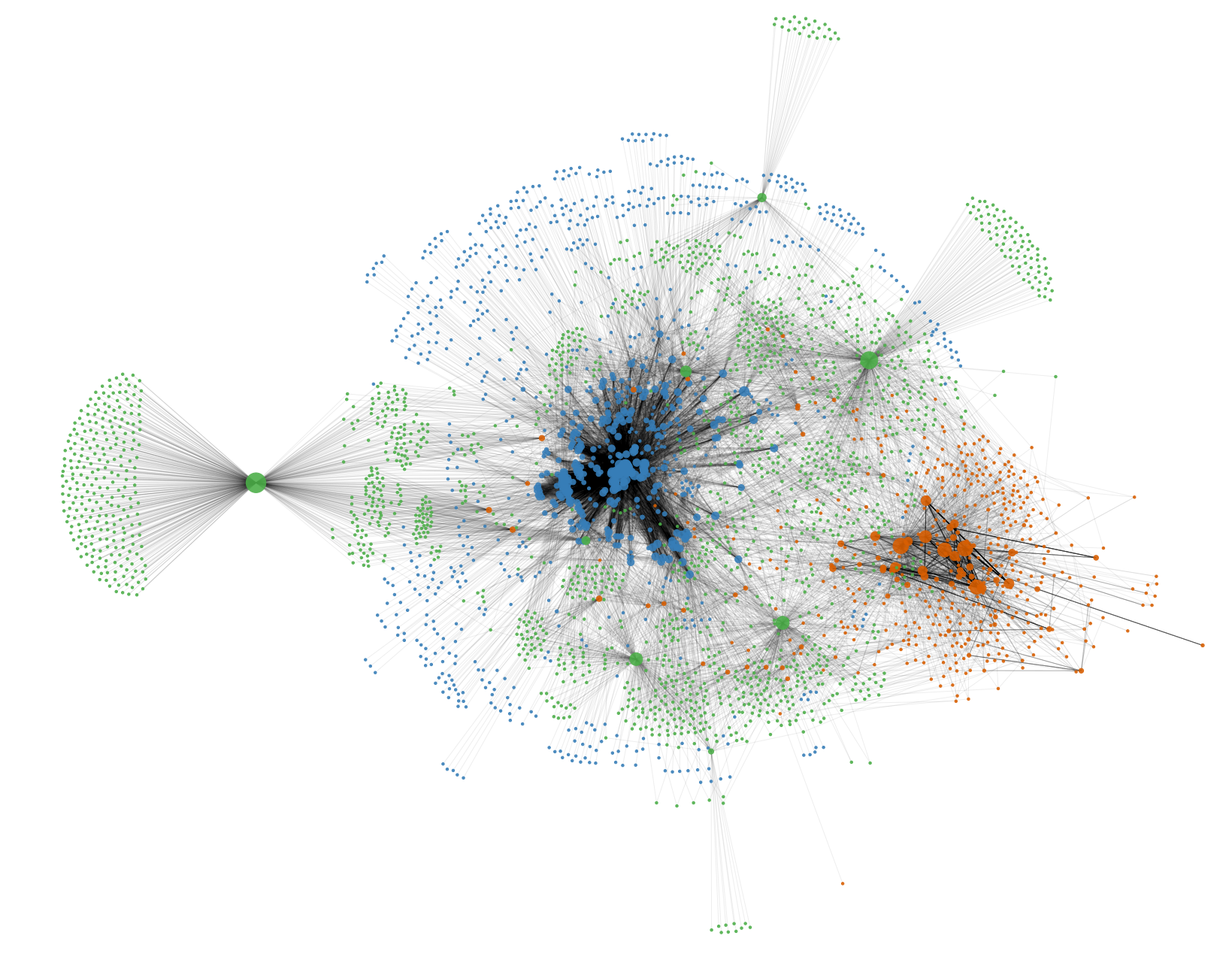}
         \caption{Example DKG constructed from ontology with no labels, but color coding set to match ontology.}
        \label{fig:kg_baltimore}
     \end{subfigure}
     \hspace{1em}
     \begin{subfigure}[t]{.31\textwidth}
         \centering
         \includegraphics[width=0.99\textwidth]{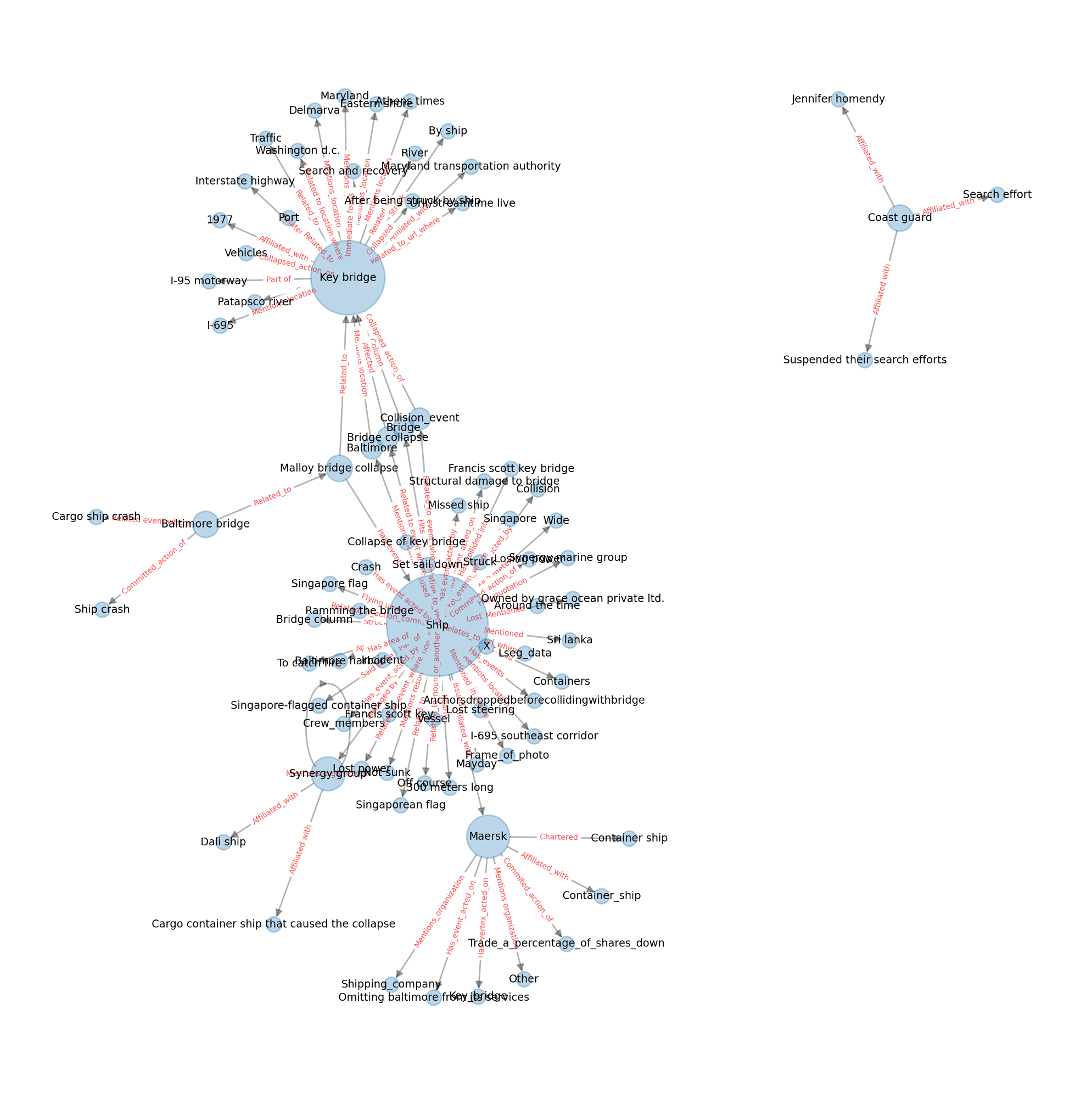}
         \caption{Example LKG constructed corpus of text using LlamaIndex.}
         \label{fig:GDELT_LI_KG}
     \end{subfigure}
     \begin{subfigure}[t]{.31\textwidth}
         \centering
         \includegraphics[width=0.99\textwidth]{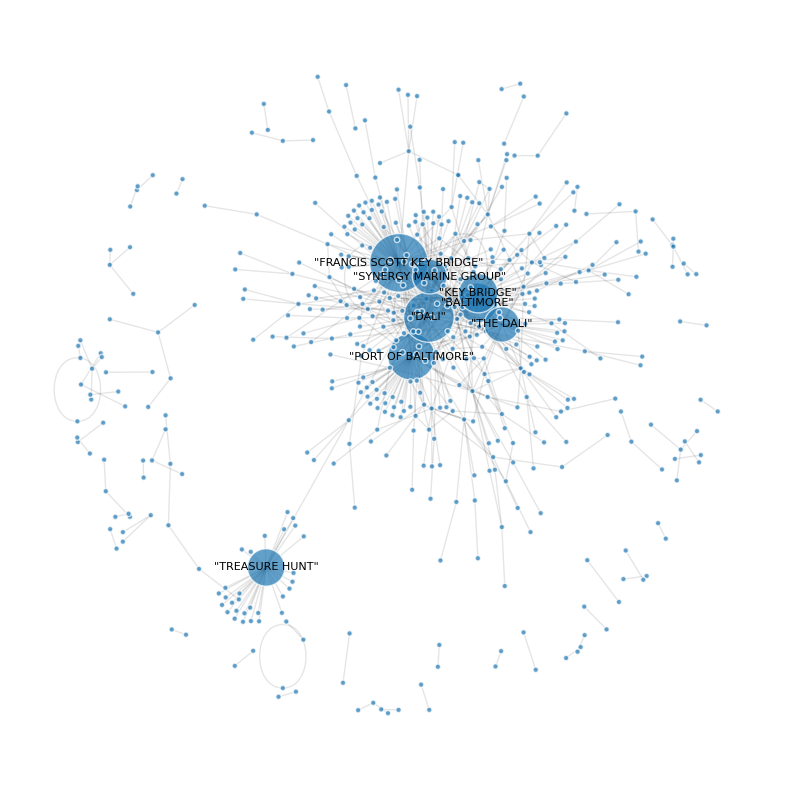}
         \caption{Example GRKG constructed corpus of text using GraphRAG, removing all isolated nodes. Large nodes have degree $\ge25$.}
         \label{fig:GDELT_GR_KG}
     \end{subfigure}
     \label{fig:KGs_from_GDELT}
     \caption{KG formations from GDELT Data of Baltimore Bridge collapse event. This subset of data included 27 articles with 283 related mentions to 143 events during the Baltimore bridge collapse from midnight to 10:00 AM EST. The corpus of text was created from web scraping the 27 URLs associated to the articles.}
     \label{fig:KGs_from_GDELT}
\end{figure}

Using the GDELT data directly and the scraped text we can construct a total of three KGs:

\begin{description}
    
\item[Direct KG (DKG):] The first KG was simply a direct conversion of the subset of the original GDELT data into an equivalent KG as specified by our ontology in Fig.~\ref{fig:gdelt2_ontology_draft}. This KG is shown in Fig.~\ref{fig:kg_baltimore}.

\item[LlamaIndex KG (LKG):] The second KG was generated by an LLM deployed against a  text corpus consisting of the source articles collected by scraping URLs of the 209 articles included in our GDELT subset, and enriched with knowledge of the ontology. This KG is shown in Fig.~\ref{fig:GDELT_LI_KG}.

\item[GraphRAG KG (GRKG):] The final KG was generated using the same articles as the LKG, using Microsoft's open-source GraphRAG package with default configuration parameters. This KG is shown in Fig.~\ref{fig:GDELT_GR_KG}.

\end{description}

The example KG constructed using our ontology (DKG) is shown in a reduced form in Fig.~\ref{fig:kg_baltimore}. The nodes are color coded based on their source. Note that node and edge labels are not shown in order to facilitate clarity. This KG is one component and has 3,469 nodes and 18,052 edges.

To construct a KG directly from the corpus of source document text (LKG) shown in Fig.~\ref{fig:GDELT_LI_KG}, we used Mixtral-8x7B~\cite{jiang2024mixtralexperts} as our base model, following the procedure outlined by the LlamaIndex package developers \cite{LLamaIdxKG}. The LLM is prompted to extract triples from the news articles according to a prompt provided in the parameter \texttt{kg\_triplet\_prompt}. Using the default prompt, the ontology does not get incorporated and the resulting KG is a star-shaped graph with a single central node and all other nodes connected to this center, there being no other edges.  When we change the prompt to consider the entire ontology, we again get a star-shaped graph. Nontrivial graph structure arose when we prompted the language model with a reduced version of the ontology with adaptation for unstructured text. In particular, our prompt asked for: 
\begin{itemize}
    \item Vertices of one of the following types: ``Event'', ``Article'', ``Mention'', ``Person'', ``Quotation'', ``Organization'', ``Location'', and ``Other'', with the last type serving as a catch-all. 
    \item Edges of one  of the following type: ``Related to event where'', ``Has quotation'', ``Mentions person'', ``Mentions location'', ``Mentions organization'', ``Committed action of'', ``Has event acted on'', and ``Affiliated with'', again with the last type serving as a catch-all. 
\end{itemize}
While this reduced description produces a non-trivial knowledge graph, it is worth noting that node and edge types still struggle to adhere to the prescribed structure, potentially due to the nature of hallucinations. The results are shown in Fig.~\ref{fig:GDELT_LI_KG}. This pipeline also struggles with entity and relation resolutions, for example creating separate nodes for `Container ship' and `Container\_ship.'

The construction of the GRKG required a language model whose context window exceeded the capabilities of Mixtral-8x7B. We decided to use Llama-3.1-8B~\cite{dubey2024llama3herdmodels} for this case. It is worth noting that GraphRAG indiscriminately generates relations without the prescription of any ontology. It does, however, identify a specified set of entities with defaults being ``organizations'', ``persons'', ``geo(locations)'', and ``events.'' Similar to the LlamaIndex pipeline, GraphRAG struggles with entity resolution --- an example of which can be seen from the existence of separate nodes for ``DALI'' and ``THE DALI.'' It also spawns many small components, often being isolated nodes; 435 of 968 total nodes are isolated in this example.

\subsection{Knowledge Graph Analysis Methodology} \label{sec:kgmeth}

The three graphs in  Fig.\ \ref{fig:KGs_from_GDELT} show the significant size difference between the DKG, LKG, and GRKG. This is potentially due to the summarizing nature of LlamaIndex and GraphRAG to only capture the key edges and nodes. Further, we find that the LLM used to create the LKG had considerable trouble with adhering to the prescribed ontology, creating many new edge types outside those originally prescribed. For example, the LLM creates a triple \texttt{(Maersk, Chartered, Container ship)} when using LlamaIndex. While factually correct, the \texttt{Chartered} edge type is not specified in the LLM prompt. Even though the LLM struggles to adhere to our proposed structures, many of the edges that are mined from the article text are easily interpretable.

To validate the quality of our ontology and to show some of its use cases in an automatic way we provide a qualitative comparison where we use an LLM for question-answering on the GDELT dataset. 
In total we have five pipelines on how we implement an LLM for talking to GDELT. These pipelines are shown in Fig.~\ref{fig:experiments_pipeline}, where in each case, we use an LLM to produce a final answer from the information obtained from each retrieval method. We note that this final processing is done with models with 7-8B parameters. With the GraphRAG pipeline we use Llama-3-8B~\cite{dubey2024llama3herdmodels} for question answering, and in all other cases we use Mistral-7B~\cite{jiang2023mistral7b} in tandem with the E5-large-v2~\cite{wang2022text} embedding model. In order, left to right:

\begin{enumerate}
\item Extract knowledge using direct graph queries to probe the DKG. This method requires an analyst to convert a natural language question into a suitable graph query. Consistent with other techniques, we then use an LLM to interpret and repackage the retrieved information.
\item Use G-retriever to automatically fetch a subgraph of the GDELT Knowledge Graph from a natural language question. This subgraph retrieval algorithm requires vectorizing the nodes and edges of the knowledge graph using a language embedding model. Once retrieved, the subgraph is fed into an LLM for interpretation.
\item Construct a knowledge graph by parsing full-text documents (scraped from GDELT URLs) with an LLM and LlamaIndex's functionality. We then proceed as in (2).
\item Create a Vector Store using the text Corpus and ask questions in a typical RAG setting. This involves using a language embedding model to vectorize text articles (after splitting them into 500-token chunks). Given a question, we use the same embedding model to vectorize it and, using the Euclidean metric to identify nearby vectors, extract the most similar text snippets. The original question and its associated context are then fed into an LLM to process an answer.
\item Build a knowledge graph as in (3), using the GraphRAG ecosystem. We also use the provided question-answering capabilities of the GraphRAG package.
\end{enumerate}

We note that it is typically unnecessary to use an LLM in method (1) and the answer can often be inferred after seeing the raw output of the graph query. In this case, the LLM effectively repackages the result of the graph query in an expected way. As such, we view this first method as a `ground-truth' on the constructed KG. Tautologically, if the DKG can be used to answer a question, then a suitable graph query is able to extract the correct answer. Conversely, if the KG cannot be used to answer a given question, then a suitable graph query can prove that the answer cannot be found in the KG. 

It is worth noting why we only employ graph queries on the DKG and not those constructed using LlamaIndex or GraphRAG. As noted in our empirical observations of the LLM-produced knowledge graphs, these graphs have far less structure than GDELT itself. This lack of defined structure makes it difficult to form useful and interesting graph queries.

\subsection{Results}

\begin{table}[h]
{\footnotesize \resizebox{\columnwidth}{!}{%
\begin{tabular}{llllll}
 & \begin{tabular}[c]{@{}l@{}}Graph Query \\ on DKG\end{tabular} & \begin{tabular}[c]{@{}l@{}}G-Retriever \\ on DKG\end{tabular} & \begin{tabular}[c]{@{}l@{}}RAG using \\ Vector Store\end{tabular} & \begin{tabular}[c]{@{}l@{}}G-Retriever \\ on LKG\end{tabular} &
 \begin{tabular}[c]{@{}l@{}}GraphRAG Q\&A \\ on GRKG\end{tabular}\\ \cline{2-6} 
\multicolumn{1}{l|}{\begin{tabular}[c]{@{}l@{}}What is the name of \\ the Bridge that \\ collapsed and what \\ river was it on?\end{tabular}} & \multicolumn{1}{l|}{\cellcolor[HTML]{CBF5CB}\begin{tabular}[c]{@{}l@{}}The Francis Scott Key\\ Bridge and it was on \\ the Patapsco River \\ in Maryland.\end{tabular}} & \multicolumn{1}{l|}{\cellcolor[HTML]{FFCCC9}\begin{tabular}[c]{@{}l@{}}The bridge is located\\ in Sri Lanka. \\ However, there is no\\ explicit mention of \\ the river's name.\end{tabular}} & \multicolumn{1}{l|}{\cellcolor[HTML]{CBF5CB}\begin{tabular}[c]{@{}l@{}}The Francis Scott Key\\ Bridge collapsed into \\ the Patapsco River.\end{tabular}} & \multicolumn{1}{l|}{\cellcolor[HTML]{FFFFC7}\begin{tabular}[c]{@{}l@{}}The bridge that \\ collapsed spanned \\ over the Patapsco \\ river.\end{tabular}} & \multicolumn{1}{l|}{\cellcolor[HTML]{CBF5CB}\begin{tabular}[c]{@{}l@{}}The Francis Scott Key\\ Bridge which spans the \\ the Patapsco River.\end{tabular}} \\ \cline{2-6} 
\multicolumn{1}{l|}{\begin{tabular}[c]{@{}l@{}}What is the name of\\ the ship that collided \\ with the baltimore \\ bridge?\end{tabular}} & \multicolumn{1}{l|}{\cellcolor[HTML]{FFFFC7}\begin{tabular}[c]{@{}l@{}}The name of the ship\\ is not provided. \\ However, the ship is a\\ container ship and \\ Singapore-flagged.\end{tabular}} & \multicolumn{1}{l|}{\cellcolor[HTML]{EFEFEF}\begin{tabular}[c]{@{}l@{}}There is no mention\\ of the name of the \\ ship that collided \\ with the bridge.\end{tabular}} & \multicolumn{1}{l|}{\cellcolor[HTML]{CBF5CB}\begin{tabular}[c]{@{}l@{}}The name of the ship \\ that collided with the \\ Baltimore bridge is \\ the Dali.\end{tabular}} & \multicolumn{1}{l|}{\cellcolor[HTML]{FFCCC9}\begin{tabular}[c]{@{}l@{}}The name of the \\ ship that collided \\ with the Baltimore \\ bridge is Maersk.\end{tabular}} & \multicolumn{1}{l|}{\cellcolor[HTML]{CBF5CB}\begin{tabular}[c]{@{}l@{}}The ship that collided \\ with the Baltimore \\ bridge is the Dali.\end{tabular}} \\ \cline{2-5} 
\multicolumn{1}{l|}{\begin{tabular}[c]{@{}l@{}}How many articles \\ did CNN publish \\ about the Baltimore \\ bridge collapse?\end{tabular}} & \multicolumn{1}{l|}{\cellcolor[HTML]{CBF5CB}\begin{tabular}[c]{@{}l@{}}CNN published 5 \\ articles about the \\ Baltimore bridge \\ collapse.\end{tabular}} & \multicolumn{1}{l|}{\cellcolor[HTML]{FFCCC9}\begin{tabular}[c]{@{}l@{}}CNN published a \\ total of 2 articles \\ about the Baltimore\\ bridge collapse.\end{tabular}} & \multicolumn{1}{l|}{\cellcolor[HTML]{FFCCC9}\begin{tabular}[c]{@{}l@{}}CNN published three \\ articles about the \\ Baltimore bridge \\ collapse.\end{tabular}} & \multicolumn{1}{l|}{\cellcolor[HTML]{FFCCC9}\begin{tabular}[c]{@{}l@{}}There is only one \\ article published by \\ CNN about the \\ Baltimore bridge \\ collapse.\end{tabular}} & \multicolumn{1}{l|}{\cellcolor[HTML]{FFFFC7}\begin{tabular}[c]{@{}l@{}}CNN published at \\ least two articles \\ about the Baltimore\\ bridge collapse.\end{tabular}} \\ \cline{2-6} 
\multicolumn{1}{l|}{\begin{tabular}[c]{@{}l@{}}On what date did \\ the Baltimore \\ Bridge collapse?\end{tabular}} & \multicolumn{1}{l|}{\cellcolor[HTML]{CBF5CB}\begin{tabular}[c]{@{}l@{}}The Baltimore Bridge \\ collapsed on \\ March 26, 2024.\end{tabular}} & \multicolumn{1}{l|}{\cellcolor[HTML]{EFEFEF}\begin{tabular}[c]{@{}l@{}}I cannot directly \\ answer that question\\ based on the given \\ data.\end{tabular}} & \multicolumn{1}{l|}{\cellcolor[HTML]{CBF5CB}\begin{tabular}[c]{@{}l@{}}The Baltimore Bridge \\ collapsed on \\ March 26, 2024.\end{tabular}} & \multicolumn{1}{l|}{\cellcolor[HTML]{FFFFC7}\begin{tabular}[c]{@{}l@{}}The Baltimore Bridge \\ collapsed at 1:20 a.m.\end{tabular}} & \multicolumn{1}{l|}{\cellcolor[HTML]{CBF5CB}\begin{tabular}[c]{@{}l@{}}The Baltimore Bridge \\ collapsed on \\ March 26, 2024.\end{tabular}} \\ \cline{2-6} 
\multicolumn{1}{l|}{\begin{tabular}[c]{@{}l@{}}Who is \\ Brandon Scott?\end{tabular}} & \multicolumn{1}{l|}{\cellcolor[HTML]{FFFFC7}\begin{tabular}[c]{@{}l@{}}Brandon Scott is a \\ person mentioned in \\ several news articles \\ related to the collapse \\ of the Francis Scott \\ Key Bridge.\end{tabular}} & \multicolumn{1}{l|}{\cellcolor[HTML]{CBF5CB}\begin{tabular}[c]{@{}l@{}}Brandon Scott is \\ the Mayor of \\ Baltimore.\end{tabular}} & \multicolumn{1}{l|}{\cellcolor[HTML]{CBF5CB}\begin{tabular}[c]{@{}l@{}}Brandon Scott is \\ the mayor of \\ Baltimore, Maryland.\end{tabular}} & \multicolumn{1}{l|}{\cellcolor[HTML]{EFEFEF}\begin{tabular}[c]{@{}l@{}}Brandon Scott is not \\ mentioned in the \\ given data.\end{tabular}} & \multicolumn{1}{l|}{\cellcolor[HTML]{CBF5CB}\begin{tabular}[c]{@{}l@{}}Brandon Scott is \\ the mayor of \\ Baltimore, Maryland.\end{tabular}} \\ \cline{2-6}  
\multicolumn{1}{l|}{\begin{tabular}[c]{@{}l@{}}Who is \\ Niki Fennoy?\end{tabular}} & \multicolumn{1}{l|}{\cellcolor[HTML]{FFFFC7}\begin{tabular}[c]{@{}l@{}}Niki Fennoy is a \\ person mentioned in \\ various news articles \\ related to the collapse \\ of the Francis Scott \\ Key Bridge.\end{tabular}} & \multicolumn{1}{l|}{\cellcolor[HTML]{FFFFC7}\begin{tabular}[c]{@{}l@{}}The article from \\ \texttt{thepeninsular}-\\-\texttt{qatar.com} \\ mentions Niki\\ Fennoy.\end{tabular}} & \multicolumn{1}{l|}{\cellcolor[HTML]{EFEFEF}\begin{tabular}[c]{@{}l@{}}I don't know. Niki \\ Fennoy is not menti-\\ -oned in the provided\\ context.\end{tabular}} & \multicolumn{1}{l|}{\cellcolor[HTML]{EFEFEF}\begin{tabular}[c]{@{}l@{}}Niki Fennoy is not \\ present in the \\ given data.\end{tabular}} & \multicolumn{1}{l|}{\cellcolor[HTML]{CBF5CB}\begin{tabular}[c]{@{}l@{}}Niki Fennoy is a \\city police spokesman.\end{tabular}} \\ \cline{2-6} 
\multicolumn{1}{l|}{\begin{tabular}[c]{@{}l@{}}What are the top\\ themes present in\\ the data?\end{tabular}} & \multicolumn{1}{l|}{\cellcolor[HTML]{CBF5CB}\begin{tabular}[c]{@{}l@{}}\tiny{MARITIME\_INCIDENT} \\ \tiny{MARITIME} \\ \tiny{MANMADE\_DISASTER} \\ \tiny{TAX\_FNCACT} \\ \tiny{WB\_137\_WATER}.\end{tabular}} & \multicolumn{1}{l|}{\cellcolor[HTML]{FFFFC7}\begin{tabular}[c]{@{}l@{}}\tiny{MARITIME\_INCIDENT} \\\tiny{CRISIS} \\ \tiny{TAX}\\\tiny{NEWS} \\\tiny{ETHNICITY}.\end{tabular}} & \multicolumn{1}{l|}{\cellcolor[HTML]{EFEFEF}\begin{tabular}[c]{@{}l@{}}I don't have enough \\information from \\to determine specific \\the themes present\\in the data.\end{tabular}} & \multicolumn{1}{l|}{\cellcolor[HTML]{FFCCC9}\begin{tabular}[c]{@{}l@{}}\tiny{EVENTS AND THEIR} \\\tiny{RELATIONSHIPS},\\ \tiny{LOCATIONS},\\ \tiny{ORGANIZATIONS}, \\ \tiny{VESSELS}.\end{tabular}} & \multicolumn{1}{l|}{\cellcolor[HTML]{FFFFC7}\begin{tabular}[c]{@{}l@{}}\tiny{NEWS AND UPDATES} \\\tiny{BRIDGE COLLAPSE} \\ \tiny{CONSTRUCTION CREW}\\\tiny{SEARCH AND RESCUE} \\\tiny{COMMUNITY REPORT}.\end{tabular}} \\ \cline{2-6}
\end{tabular}}} 
\caption{Table of example questions and answers highlighting deficiencies in each method for analyzing the GDELT data. Table highlight color legend: Green is a correct answer, yellow is a partially correct answer, red is an incorrect answer, and grey is for no answer provided.}
\label{tab:questions}
\end{table}

Table \ref{tab:questions} shows a sample of questions that were passed through each of the five pipelines from Fig.~\ref{fig:experiments_pipeline}. Exact queries to the GDELT knowledge graph were generated by searching for keywords in the edge triples comprising the knowledge graph. Specifically, we searched for keywords in these triples by converting each triple to a sentence (stored as a string) and searching therein. We then used the edge induced subgraph from the edge sentences where keywords were found. The following keywords were used for each question:
{\small
\begin{itemize}
    \item What is the name of the Bridge that collapsed and what river was it on?: \textbf{Bridge, Collapse, River}
\item What is the name of the ship that collided with the baltimore bridge?: \textbf{Ship, Collide, Baltimore, Bridge}
\item How many articles did CNN published about the baltimore bridge collapse?: \textbf{CNN, Baltimore, Bridge, Collapse}
\item On what date did the Baltimore Bridge collapse?: \textbf{Date, Baltimore, Bridge, Collapse}
\item Who is Brandon Scott?: \textbf{Brandon Scott}
\item Who is Niki Fennoy?: \textbf{Niki Fennoy}
\item What are the top themes present in the data?: \textbf{Has\_Theme}
\end{itemize}
}
\noindent Our prompt was then constructed as the following: ``Please answer the question given the following information:" with the list of edge sentences appended to the end.

Solutions based on vector stores, GraphRAG, and direct graph queries on the DKG offer the best results for question answering. With direct graph queries, we can answer high-level and other types of questions that need us to aggregate information across the dataset. For example, we can easily extract information about recurrent themes or about a particular news source. While GraphRAG also provides functionality to answer high-level questions, we found that its performance lacked in this respect. It performed well on fine-grained questions. A vector store performed similarly well on these fine-grained questions that can be answered by a small number of excerpts from the source articles. Notably, the second column suggests that the automated search functionalities provided in G-retriever are often unable to retrieve a subgraph that can be used to provide an accurate answer. The problem gets worse when we use the KG created by LlamaIndex as our knowledge base, which is unable to suitably answer any question that we posed. These issues may partially be due to the out-of-box application of the G-retriever system and careful fine-tuning may improve performance. Regardless, we see the retained value in directly probing our data with hand-crafted queries and infer that further developments are needed for these automated information extraction systems to match the baseline performance on questions that require reasoning across the entire corpus.

By combining the results of the direct graph queries on the DKG with those of GraphRAG and standard RAG, we can provide suitable answers to the all of the presented questions. On one hand, we see that the DKG can provide better responses to high-level or aggregate questions about our data. Conversely, the vector store and GRKG can be used to identify local information in the documents that might be missing in the DKG.

Regarding popular subjects, direct prompting of the LLM without dealing with knowledge graphs or vector stores reveals that the LLM independently recognizes Brandon Scott as the mayor of Baltimore. For all other questions, the language model cannot answer the posed questions by itself due to the recency of the bridge collapse.

To quantitatively evaluate the quality of the answers generated by our different question-answering methods, we require a set of ground truth answers for the questions posed. Table \ref{tab:ground_truth} presents these manually curated ground truth answers, representing the expected correct responses for each question used in our evaluation. These ground truth answers serve as the benchmark against which the performance of each system is measured.

\begin{table}[h]
{\footnotesize \resizebox{\columnwidth}{!}{%
\begin{tabular}{p{0.4\columnwidth} p{0.5\columnwidth}}
\textbf{Question} & \textbf{Ground Truth} \\ \hline
What is the name of the Bridge that collapsed and what river was it on? & The Francis Scott Key Bridge on the Patapsco River. \\ \hline
What is the name of the ship that collided with the baltimore bridge? & The ship was named the Dali. \\ \hline
How many articles did CNN publish about the Baltimore bridge collapse? & CNN published 5 articles. \\ \hline
On what date did the Baltimore Bridge collapse? & The collapse occurred on March 26, 2024. \\ \hline
Who is Brandon Scott? & Brandon Scott is the Mayor of Baltimore. \\ \hline
Who is Niki Fennoy? & Niki Fennoy is a city police spokesman. \\ \hline
What are the top themes present in the data? & Themes include maritime incidents, manmade disaster, and water-related topics. \\
\end{tabular}
}}
\caption{Ground Truth Answers for the Baltimore Bridge Collapse Questions}
\label{tab:ground_truth}
\end{table}

\begin{figure}[h!]
    \centering
    \includegraphics[width=0.75\textwidth]{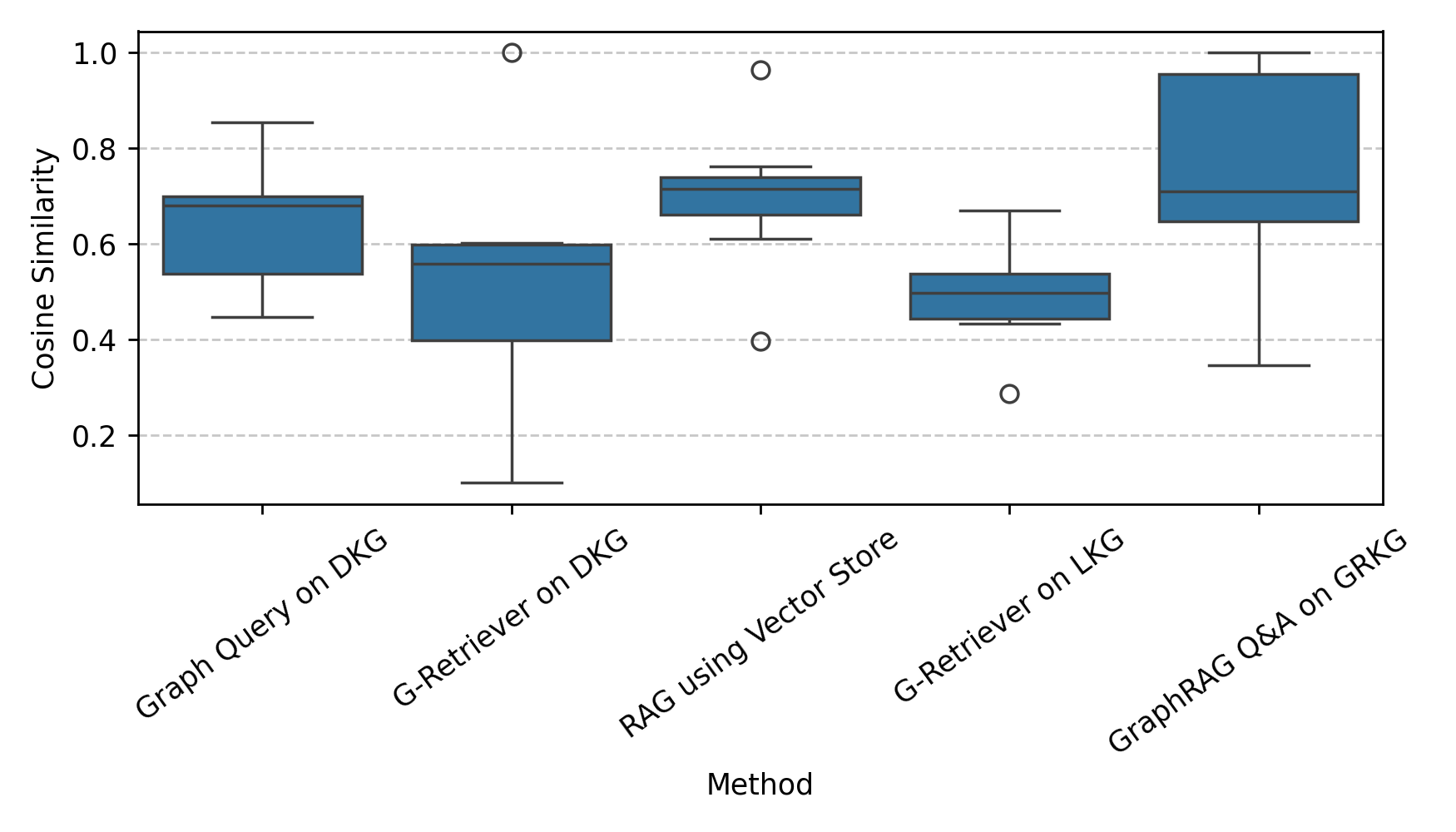}
    \caption{Box plots comparing the cosine similarity scores of different question-answering methods applied to the GDELT data concerning the Baltimore bridge collapse. Higher cosine similarity indicates a greater semantic similarity between the predicted and actual answers.}
    \label{fig:cos_sim_gdelt}
\end{figure}

To further validate our qualitative findings, Figure \ref{fig:cos_sim_gdelt} presents a quantitative comparison of the semantic similarity of answers generated by the five different methods when querying the GDELT dataset related to the Baltimore bridge collapse. We compute the quality of the result by calculating cosine similarity\footnote{The semantic embeddings used to calculate the cosine similarity were generated using the \texttt{sentence-transformers/all-MiniLM-L6-v2} model from the Sentence Transformers library.}, a measure of the similarity between the embeddings of the predicted answer and the ground truth, with higher values indicating greater accuracy. The box plots illustrate the distribution of these similarity scores for each method: Graph Query on DKG, G-Retriever on DKG, RAG using Vector Store, G-Retriever on LKG, and GraphRAG Q\&A on GRKG. As our qualitative analysis suggested, methods leveraging direct graph queries on the DKG, standard RAG with a vector store, and GraphRAG Q\&A on the GRKG generally exhibit the highest cosine similarity scores, providing quantitative evidence for their superior performance in generating accurate and relevant answers compared to the G-Retriever, especially with the LKG. This visualization quantitatively confirms the trends observed in the qualitative evaluation presented in Table \ref{tab:questions}.

\section{Conclusion}

This work has taken the GDELT GKG database and introduced an ontology to create a knowledge graph with rich structure. 
We found that while the large KG produced by the GDELT-GKG2 database has rich information for question-answering, the {\it ad hoc} techniques for graph exploration deem further investigation for reliable use. That said, the information stored in the KG produced here is not without flaws, and we expect the example studied here to be useful for the broader synergy between KGs and LLMs as addressed by others \cite{Pan_2024}. While the KGs produced using LlamaIndex captured a summary of the events surrounding the Baltimore bridge collapse, the quality of the resulting structure was not suitable for question-answering. Those techniques incorporated into the GraphRAG package did considerably better, but there is room for improvement for answering global questions, resolving duplicate entities, and incorporating ontologically-guided relation extraction. We believe the incorporation of the ontology into the relation extraction will have the benefit of allowing tools such as GraphRAG to better answer quantitative questions that only the ontology based KG pipeline (DKG) was able to answer (e.g., the number of articles published on a topic).

%\TGComment{Let's think hard if the sentences after this in this paragraph are saying what we want.  GraphRAG does pretty well on global stuff.  DKG comparisons are pretty unfair}
Large language models continue to be adapted to solve problems across domains, and the case study on the KG built here presents much opportunity for future development. In particular, the debate around using raw documents or a curated KG should not be of `Either/Or', but rather integration between these two modalities. Our findings demonstrate the significant value of combining the strengths of both approaches. We believe that by combining news articles in the form of a vector store with the rich ontological structure of the GDELT knowledge graph through an LLM interface, the resulting information extraction system would allow for better knowledge retrieval than either component alone. 
Specifically, the direct knowledge graph (DKG) excels at answering high-level or aggregate questions, providing a strong foundation of structured knowledge. 
While the vector store is well-suited for identifying local information within the documents that might be missing or less readily accessible in the DKG, GraphRAG has shown effectiveness on both local and global information. 
Therefore, we hypothesize that the optimal approach is the integrated system, leveraging the DKG for broad context and the vector store and GraphRAG for detailed, document-specific insights, and for enhanced global information retrieval.
Further work must be done to determine the extent to which the textual article data can be used to refine the KG produced here; some of our examples showed that information stored in the DKG does not truthfully reflect the information in the articles. Conversely, we must determine how the constructed KG can be used to better search the associated vector store. Some of these research directions include the following:
\begin{itemize}
    \item Use LLMs to add new information to an existing KG by creating new entities or edge relations. Based off our observations with LlamaIndex and GraphRAG, we need careful monitoring to ensure that the LLM formats its responses to properly adhere to the ontological structures and other existing structures in the KG. To this end, it can be beneficial to adapt the triples in the DKG produced here to fine-tune the language model or guide its output with in-context learning.
    \item By introducing RAG capabilities, we can fact check the KG against the raw textual information. For example, we found that Niki Fennoy was only mentioned in 3 articles but misattributed to 11 others. The use of LLMs give a potential avenue to fact-check existing relations. 
\end{itemize}

\section{Acknowledgements}
This work is under information release number PNNL-SA-209193.

\bibliographystyle{plain}
\bibliography{refs}

\end{document}